# Astronomical Evidence for the Rapid Growth of Millimeter Sized Particles in Protoplanetary Disks

Jonathan P. Williams
Institute for Astronomy, University of Hawaii, Honolulu, USA

ABSTRACT

I summarize recent surveys of protoplanetary disks at millimeter wavelengths and show that the distribution of luminosity, equivalent to the mass in small dust grains, declines rapidly. This contrasts with statistics on the lifetime of disks from infrared observations and the high occurrence of planets from radial velocity and transit surveys. I suggest that these disparate results can be reconciled if most of the dust in a disk is locked up in millimeter and larger sized particles within about 2 Myr. This statistical result on disk evolution agrees with detailed modeling of a small number of individual disks and with cosmochemical measurements of rapid planetesimal formation.

## 1. INTRODUCTION

Almost all sun-like stars are born surrounded by disks of gas and dust. These disks are a direct and inevitable consequence of angular momentum conservation, as molecular cloud cores gravitationally collapse to stellar scales, and they are the birth sites of planets. The recent discoveries of the *Kepler* mission and sensitive radial velocity surveys show that planets are common, more the norm than not, with a number distribution that strongly increases as planet size and mass decrease (Youdin 2011; Howard et al. 2010). In short, disks are almost ubiquitous, relatively long-lived, and their typical end state is a planetary system. Here I briefly summarize millimeter wavelength surveys of protoplanetary disks and how they can inform us about the first steps of planetesimal formation. As the focus of this journal is on cosmochemistry, I include some basic information on astronomical observations of dust and try to draw some connections to meteoritic studies.

## 2. DISK LIFETIMES INFERRED FROM INFRARED SURVEYS

Protoplanetary disks are brightest at infrared wavelengths (~1-100μm). The cool dust emits most strongly at these wavelengths whereas the star peaks in the optical and becomes fainter in the infrared. Because the disk is much larger in area than the star, it therefore typically dominates the combined emission beyond about 2µm. Ground-based observations are efficient up to about this wavelength but are strongly limited beyond it by the high instrumental and atmospheric background. With the *Infrared Astronomical Satellite (IRAS)*, *Infrared Space Observatory (ISO)*, and most recently *Spitzer* space telescopes, astronomers have been able to carry out sensitive large surveys of disks in many regions of different ages and thereby measure their statistical lifetime and determine their varied evolutionary pathways (Williams and Cieza 2011 and references therein).

These infrared surveys show that disks are very common initially, found around about 90% of young stars with ages < 1 Myr, but their occurrence decreases with time such that <5% of stars have disks in clusters with ages > 5 Myr. This result has been well established for over a decade now (Haisch, Lada, & Lada 2001). Figure 1 shows a recent compilation of *Spitzer* data and a weighted least squares exponential fit. The data are well fit by an exponential and the implied disk half-life is 2 Myr.

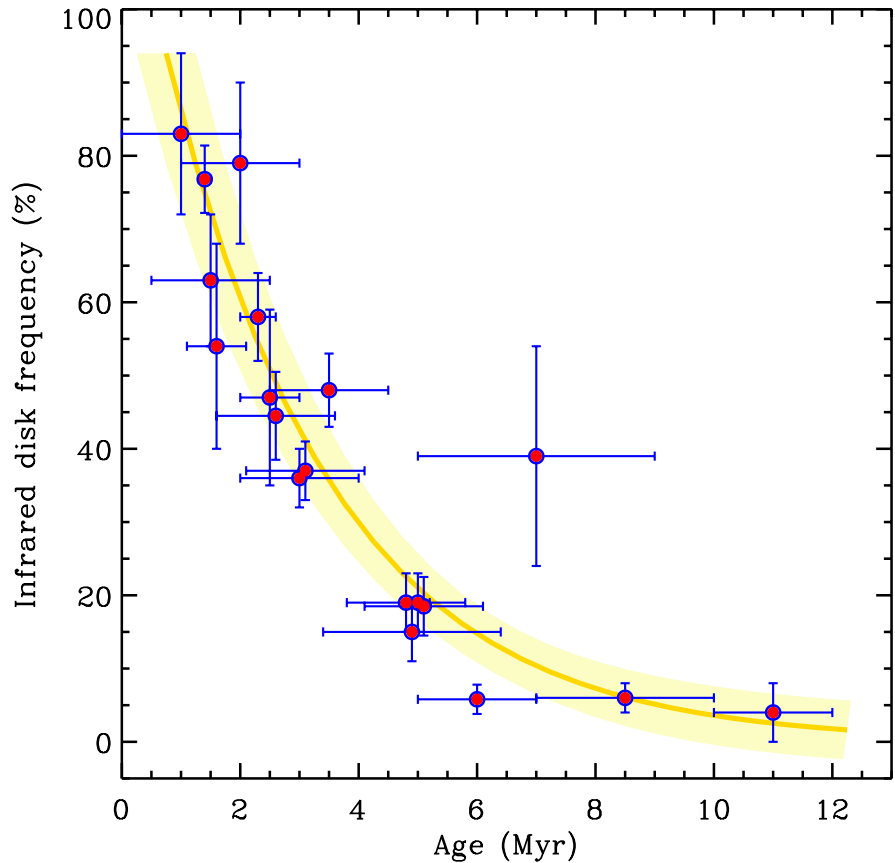

**Figure 1:** The detection rate of circumstellar disks around stars in clusters and star forming regions of different ages, as determined from short wavelength (8-24µm) infrared observations from the *Spitzer* space telescope. This plot is based on that in Hernandez et al. (2007), and updated with additional data sent to the author by Jesus Hernandez. Most stars have infrared excesses indicative of disks at early times but very few show such excesses beyond about 5 Myr. The yellow line shows a weighted least squares exponential fit to the data and has an e-folding time of 2.8 Myr, corresponding to a disk half-life of 2 Myr.

This plot is attractive in its simplicity but requires care in its interpretation. Each point is a summary of observations of tens to hundreds of young stellar objects in a single star-forming region. It is difficult to determine the age of any individual star (beside our own) with precision better than about 0.5-1 Myr because of uncertainties in protostellar evolutionary models and the distance to the star, but it is easier to determine the average age of a collection of stars that formed together particularly for nearby so-called moving groups whose expansion can be back-tracked to a common origin in time and space (de Zeeuw et al. 1999). Figure 1 plots the average age for each group and its estimated dispersion, which includes uncertainties in the average age estimate and a finite period over which the stars formed. The dispersion in the disk frequency is Poisson counting statistics. There can also be subtleties in the detection limit and sample selection for each star-forming region. In this case, however, the *Spitzer* observations are very sensitive and of relatively nearby, well characterized regions. These data are complete (i.e., able to detect disks that extend close to the stellar photosphere) down to low stellar luminosities, typically corresponding to stellar masses < 0.5 $M_\odot$.

There follows, of course, the interpretation of what this means for planet formation. From a purely astrophysical perspective, we have to consider what are we detecting and where are we seeing it. As we are considering continuum (not spectral line) observations here, we are detecting solids, i.e., dust. A general rule of thumb is that observations of dust particles at a wavelength $\lambda$ are most sensitive to grain sizes with radii $a \sim \lambda$. The reason is that particles emit very inefficiently at wavelengths larger than their size, and that particles much larger than the observing wavelength have a relatively small ratio of area to mass. Near-infrared wavelength observations therefore tell us mainly about micron-sized dust grains. This is on the high end of the size spectrum of dust grains in the diffuse interstellar medium but is not what we think of as planetesimals.

In regard to where are we seeing the dust, Wien's law tells us that their characteristic temperature is $T \sim 3000$ K/$\lambda(\mu m)$. Because disks are hot near the star and cooler further away, we can effectively map wavelength to temperature and thence to radius to learn about disk structure, at least at small radii (Dullemond et al. 2007). However, for a disk around a solar mass protostar star, most of the dust beyond $R \sim 1$ AU is too cool to emit at $\lambda < 10\mu m$. Furthermore, the short wavelength emission from the interior of the disk is absorbed by dust in the upper parts and is completely obscured from optical through far infrared wavelengths. The astronomical term to quantify this is optical depth, $\tau$, and the emission is *optically thick* when $\tau > 1$. The optical depth through a disk is $\tau = \kappa \Sigma$, where $\Sigma$ is the integrated surface area (units g cm$^{-2}$) and the dust opacity, $\kappa$ (units cm$^2$ g$^{-1}$) is the extinction (absorption plus scattering) surface area per unit mass. Theoretically, $\kappa$ is derived from Mie theory and depends on the size, shape, and refractive index of the grains. Tabulations for different mineralogical and ice mantle compositions can be found in Pollack et al. (1994) and Ossenkopf & Henning (1994).

The dust opacity in the mid-infrared, $\lambda \sim 3$-$30\mu m$ away from silicate features, is $\kappa \sim 1$-$10$ cm$^2$ g$^{-1}$, to within about an order of magnitude depending mainly on the grain size distribution. Only the inner few AU of a disk is hot enough to radiate at these wavelengths. A Minimum Mass Solar Nebula (MMSN; Weidenschilling 1977) disk has $\Sigma_{dust} \gg 1$ g cm$^{-2}$ at such radii. Thus, the emission is very optically thick, τ >> 1, and we see only the upper surface of the disk at short wavelengths.

As we proceed to longer wavelengths, we see emission predominantly from cooler and larger grains. The lower temperatures correspond to greater radii. Because bigger grains have a smaller surface area per unit mass and there are proportionally fewer than small grains, the opacity decreases with wavelength. Thus we see deeper toward the planet forming midplane of the disk.

The situation is usefully summarized in Figure 2, which shows the regions that emit half of the emission from a disk (the central 70% in radial and central 70% in vertical extent) for different λ. Scattered starlight from the surface of the disk is seen at 1-3μm, and thermal dust emission is detectable at longer wavelengths. The inner regions are hotter and emit at shorter wavelengths but the emission is optically thick so only the upper parts of the disk are seen. Longer wavelength emission comes predominantly from the cooler, outer regions of the disk but also from closer to the midplane.

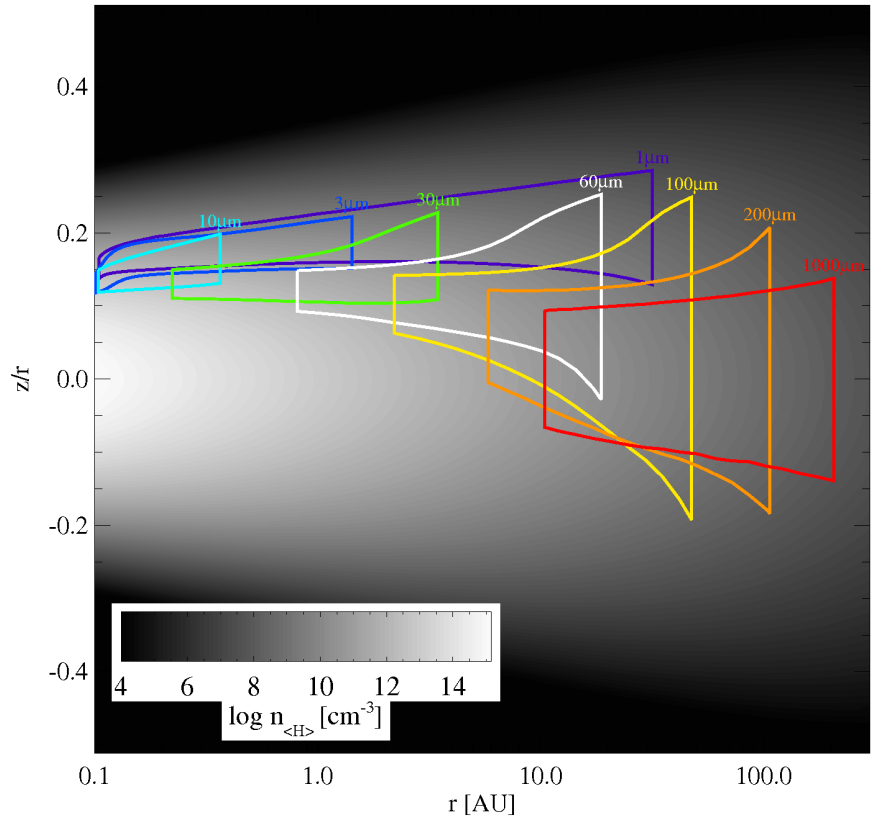

**Figure 2:** The dependence on radius and scale height of the emission at different wavelengths for a MMSN protoplanetary disk (M=0.01M$_\odot$, R=100 AU, gas-to-dust ratio=100). The grayscale shows the number density of Hydrogen atoms in the disk, n(H), in logarithmic scale. Each colored contour shows, at the labeled wavelength, the radial range and vertical extent that encompasses 15-85% of the cumulative total emission. The total area within the contour therefore accounts for 0.7x0.7=0.49 of the emission from the disk at that wavelength. Note that the axes are logarithmic in radius and linearly proportional in scale height. Figure courtesy of Peter Woitke and based on the models in Woitke et al. (2010).

The calculation in Figure 2 is made for a MMSN disk with M=0.01 M$_\odot$, R=100 AU, and a gas-to-dust ratio of 100. The same reasoning applies, and the situation is similar, for disks with different masses, radii, and scale heights. Consequently, the infrared lifetime plot in Figure 1, which is determined from observations at λ~8-24μm, only shows the presence of hot dust from the upper regions of the inner few AU of a disk.

Far-infrared observations at wavelengths ~50-300μm are emitted from radii ~1-30 AU where we expect planet formation to occur but this wavelength range is only accessible from space. At millimeter wavelengths, broadly defined here as 300μm-3mm, the emission is generally optically thin allowing us to measure the total amount of material in a disk, and therefore its *capacity* for planet formation, and also to study the growth of dust grains to sizes that are comparable to chondrules. Moreover, much of this wavelength regime is observable from the ground where new facilities promise a revolution in the sensitivity and angular resolution of observations in the coming years.

## 3. DISK MASSES INFERRED MILLIMETER WAVELENGTH OBSERVATIONS

For the reasons mentioned above, the dust opacity at millimeter wavelengths is much lower than in the infrared. Beckwith et al. (1990) provide a commonly used prescription, κ(λ)=10(λ/300μm)$^{-β}$ cm$^2$ g$^{-1}$, where the power law index, β~1-2. Figure 2 shows that most of the emission at λ~1mm comes from R>>10 AU, where we expect low surface densities, Σ$_{dust}$ << 1 g cm$^{-2}$, and therefore τ = κΣ < 1. That is, millimeter wavelength emission is optically thin and we effectively see all the emitting particles. In this case, the disk luminosity is directly proportional to the dust mass.

To convert from dust mass to total disk mass, we then multiply by the gas-to-dust ratio. This is well constrained in the ISM to be 100 (by mass) and this is the initial value in protoplanetary disks. However, as dust grains grow in a disk, they sediment to the midplane and decouple from the gas flow. Accretion along magnetic field lines onto the protostar and photoevaporation from disk surfaces will preferentially remove gas. The gas-to-dust ratio therefore evolves away from the interstellar value to nearly zero, as seen in dusty, gas poor, debris disks. For the young, protoplanetary disks described below, we use the ISM factor of 100. Spectral line observations provide a way to independently assess the gas content but the interpretation is challenging due to the complex interplay of disk structure, heating, and chemistry (Kamp et al. 2011). This is an area of active research but, for now, remains a significant and poorly understood source of uncertainty in disk mass measurements.

Over the past few years, we have surveyed several young star forming regions at millimeter wavelengths to measure the distribution of disk masses. The task is more challenging than the infrared observations summarized in Figure 1 because the emission is much weaker, but some interesting differences have become apparent. Our work began in Taurus, which is one of the nearest regions of low mass star formation and where over one hundred protostars have been identified and well characterized from the X-ray to the radio (Gudel et al. 2007). We found that protostars with little or no infrared emission had, not surprisingly, very low disk masses but infrared bright (so-called Class II) protostars were detected with a wide range of millimeter fluxes, implying a wide range of disk masses (Andrews & Williams 2005). This is due to the optically thick/thin nature of the emission at infrared/millimeter wavelengths respectively. That is, very little dust is required to produce strong infrared emission but the millimeter emission depends on the amount of dust which, these data show, can be significantly different in disks with similar infrared properties. With the important caveat of uncertainties in the dust opacity and the gas-to-dust ratio, about 15% of Taurus Class II disks have masses greater than a MMSN, $10^{-2}$ $M_{\odot}$, and about 50% have masses greater than Jupiter, $\sim 10^{-3}$ $M_{\odot}$.

We carried out similar surveys of the Ophiuchus and Orion star-forming regions (Andrews and Williams 2007; Mann and Williams 2010), each with similar median ages, $\sim 1$ Myr, to Taurus and found broadly similar results. The mass histograms are shown in the top 3 panels of Figure 3. An interesting effect was found in the massive star-forming environment of the Trapezium Cluster in Orion in that external photoevaporation erodes the weakly bound outer edges of the largest disks, resulting in a discernable lack of massive disks, >0.04 $M_{\odot}$ = 4 MMSN (Mann and Williams 2009). This effect was only seen in the central 0.3pc around the O6 star, $\theta^1$Ori C, and more massive disks such as those seen in the high mass tail of Taurus are found in the outer regions of the Trapezium Cluster.

These three regions, Taurus, Ophiuchus, and Orion, show the distribution of disk masses at early times. There had been few millimeter detections of disks in older star-forming regions, however, despite the infrared observations of disks persisting around stars for several Myr (Carpenter et al. 2005). As the instrument sensitivity has increased, we have been able to make deeper integrations and have detected a small number of disks in two older regions, IC348 at $\sim 3$ Myr and Upper Sco at $\sim 5$ Myr (Lee et al. 2011; Mathews et al. 2012). The statistics are poor but provide the first clues on the evolution of the protoplanetary disk mass function

We found that the disks in these older regions are very faint at millimeter wavelengths, and only a few were detected. Assuming the same conversion from flux to mass as for the younger systems, we conclude that there are no MMSN disks in IC348 and only one in Upper Sco. Their mass distributions, in the lower two panels of Figure 3, are clearly shifted to lower masses relative to the $\sim 1$Myr Taurus, Ophiuchus, and Orion regions. Indeed, because of the low disk luminosities, we are only able to sample the tip of the upper end of the mass distribution. The difference in maximum disk mass from $\sim 1$ Myr to >2 Myr is over an order of magnitude. Yet, as Lee et al. (2011) show, the infrared fluxes of these millimeter detected sources are very similar between Taurus and IC348. Clearly, disk evolution at millimeter wavelengths is much faster than at infrared wavelengths.

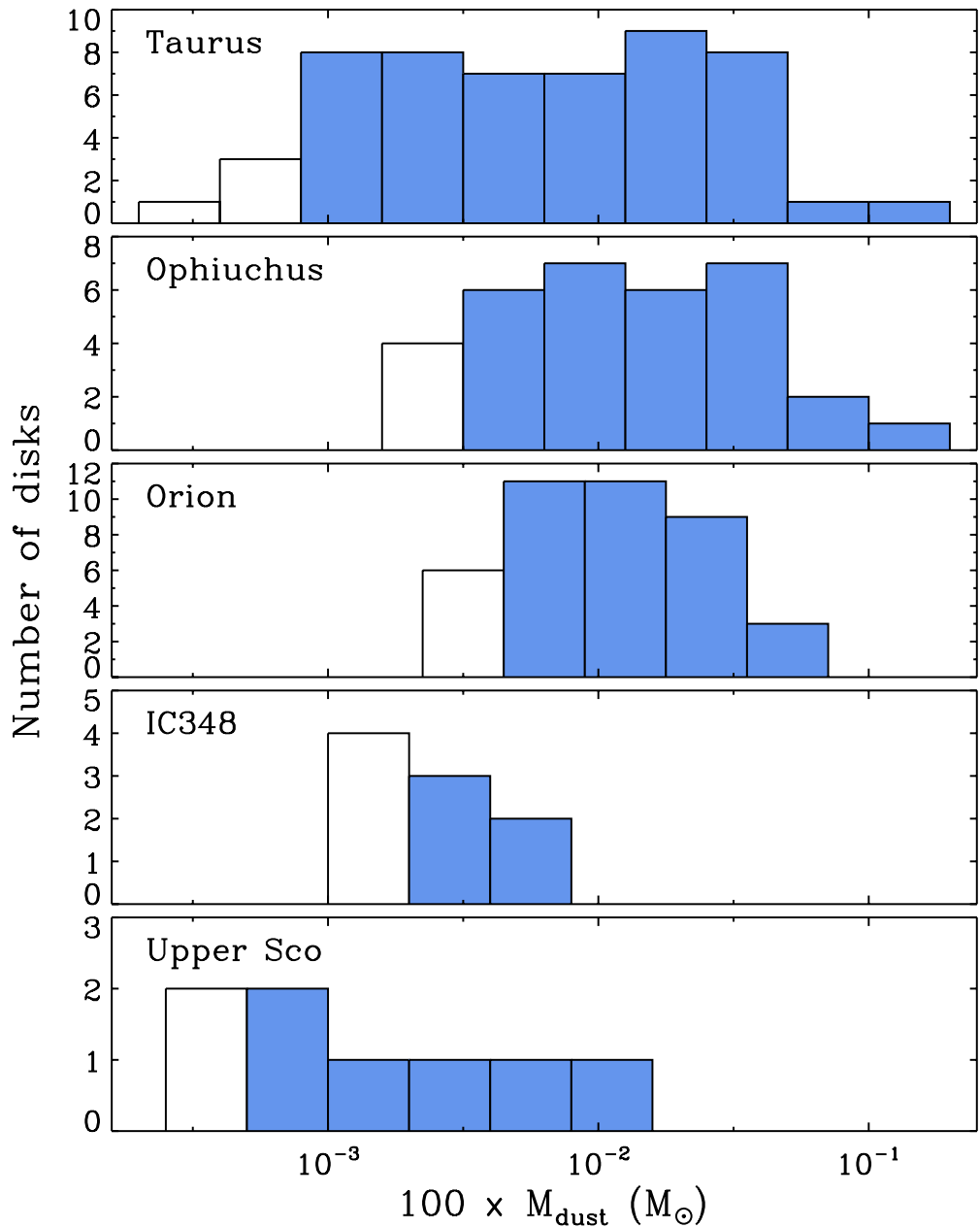

**Figure 3:** Histograms of protoplanetary disk masses in 5 regions as derived from millimeter wavelength continuum surveys. The gas-to-dust ratio of 100 on the dust mass is explicitly shown, and assumed to be the same in all disks and all regions. The blue bins shows data for which the surveys are complete, the white bins should therefore be considered lower limits to the true distribution. The top 3 panels are for young regions with median stellar ages of about 1 Myr. The lower 2 panels are older regions: 2-3 Myr for IC348 and ~5 Myr for Upper Scorpius.

## 4. INTERPRETATION

The wavelength dependent nature of disk evolution has profound implications for understanding the pace of grain growth. The astronomical study of the growth and evolution of millimeter sized particles provides interesting new avenues for comparison to cosmochemistry.

### 4.1 Rapid growth from dust to planetesimals

Very small amounts of micron-sized particles close to a star can produce the observed infrared excesses. In the millimeter wavelength studies shown in Figure 3, however, the disks become undetectable once the total mass drops below about $10^{-3}$ $M_\odot$. In terms of dust mass alone, this corresponds to $10^{-5}$ $M_\odot$, or about 3 Earth masses. Thus Figure 1 shows that *some* dust persists around *most* stars for well over 2Myr, whereas Figure 2 shows that the *amount* of small dust grains declines below the mass required for core accretion of giant planets, or for rocky super-Earths for *all* disks within this time.

In comparison, radial velocity surveys show that both Jovian mass and Earth mass planets are quite common, ~20%, in tight orbits around solar mass stars (Cumming et al. 2008; Howard et al. 2010). The statistics from the early release of Kepler transit data, which can detect planets at larger orbital radii, suggests that the median number of planets around low mass stars is greater than one (Youdin 2011). Gravitational instability of a massive disk has been invoked as an alternative, rapid formation mechanism for Jovian-type planets (Boss 1997), particularly at very large (>50 AU) orbital radii. It cannot explain the formation of dense, rocky planets, however, since the instability criteria are only met in gas-rich disks. The apparently common occurrence of Earth mass planets, that are composed mostly of metals and ices, suggests that core accretion is occurring in most protoplanetary disks.

One way to reconcile the rapid evolution of disks at millimeter wavelengths with the relatively slow timescale for rocky planet formation is for most of the dust mass to be locked up in particles much larger than a millimeter in size by about 2 Myr. That is, the decline in millimeter emission is due more to a decrease in the effective surface area, rather than the total mass of particulates. A small amount of fragmentation in grain-grain collisions suffices to explain the observed infrared excesses (Birnstiel, Klahr, and Ercolano 2012). Once grains grow beyond a millimeter in size, their dynamics decouple from the gas and a range of new mechanisms can take place such as radial drift, vertical sedimentation, and collective self-gravity (Chiang and Youdin 2010; Armitage 2010). The particles are now better thought of as planetesimals and the disk is truly protoplanetary rather than protostellar.

This statistical argument based on reconciling large multi-wavelength surveys of protoplanetary disk evolution is indirect but has the benefit of generality and tells us the timescale for the evolution of dust mass. There is other, more direct evidence for grain growth based on millimeter and longer wavelength observations alone.

## 4.2 Additional evidence for dust grain growth

Particles radiate inefficiently at wavelengths much bigger than their physical size. Consequently, a dusty object with a grain size distribution that extends only to micron sizes will radiate very inefficiently at millimeter wavelengths. This is the case for the diffuse interstellar medium where the maximum grain size is much less than a micron and which has a flux dependence on frequency, $\alpha \sim 4$, where $F_\nu \sim \nu^\alpha$, compared to $\alpha=2$ for a blackbody in the millimeter regime. However, most protoplanetary disks have a flatter spectral slope, $\alpha \sim 2$-3, at millimeter wavelengths (Andrews and Williams 2005, 2007), and beyond (Ricci et al. 2010), which indicates that the power law grain size distribution continues well beyond a millimeter and perhaps up to centimeters (Draine 2006). Furthermore, for a grain size distribution, $n(a) \sim a^{-p}$ with $p<4$, most of the dust mass would reside in these large particles.

In the particular case of the bright disk around the nearby star TW Hydrae, the thermal dust continuum slope has been detected out to 3.5cm, which indicates that nearly all the dust mass resides in centimeter and larger sized ``pebbles'', or perhaps snowballs given their cool temperature (Wilner et al. 2005).

In the near future, highly sensitive radio surveys at wavelengths from $\lambda$ = 3mm to 3cm will show the mass budget of $a \sim \lambda$ particles for a range of particle sizes in statistically significant samples of disks of different ages (Chandler and Shepherd 2008). We will then learn more about dust evolution and the timescale for growth over several important size scales that can be compared to laboratory studies of meteorites.

## 4.3 The cosmochemical connection

From a cosmochemical or planet formation perspective, the rapid pace of grain growth is not a surprise. In the inner disk, millimeter sized chondrules formed between 1-3 Myr after the first solids, CAIs (Amelin et al. 2002; Connelly et al. 2008). The parent bodies of some iron meteorites may have formed even sooner, within 1 Myr of CAI formation (Kleine et al. 2009). Furthermore, core accretion models for the formation of Jupiter and Saturn require solid cores with many Earth masses beyond 5 AU within 1 Myr (Hubickyj, Bodenheimer, & Lissauer 2005; Dodson-Robinson et al. 2008).

From the astronomical observations described here, we are beginning to learn about the amount and evolution of millimeter sized dust grains in protoplanetary disks. Ultimately, we will be able to measure the mass evolution of the solids in disks, both in aggregate and by particle size (up to about a centimeter). This can be compared to relative and absolute age measurements of meteoritic components, models of grain growth and planet formation, and opens up new avenues for cross-disciplinary research.

## 5. SUMMARY

Protoplanetary disks around young stars are most readily detected through infrared observations. The detection frequency in stellar clusters less than 1 Myr old is over 90% showing that disks are almost ubiquitous, as expected due to conservation of angular momentum. The disk frequency is below 5% in stellar clusters older than 5 Myr showing that the dust has almost all gone by then. The median disk lifetime, defined as the age range of stellar clusters with an infrared disk detection frequency of 50%, is about 2 Myr. In this article, I have reviewed surveys of protoplanetary disks in clusters of different ages at millimeter, rather than micron, wavelengths. These observations allow us to estimate the amount of dust throughout the disk, rather than simply the presence of warm dust close to the star. The data show that disks in young clusters have sufficient mass to form planets and, indeed in many cases, solar systems on the scale of our own, but that the mass declines by over an order of magnitude within the next ~2 Myr. This does not contradict the infrared observations since only a small amount of dust is required to produce detectable emission at short wavelengths. Taken at face value, this would appear to indicate that planets should be rare but the numerous radial velocity surveys and the *Kepler* results show the opposite is in fact true: planets are very common. I suggest, therefore, that the rapid evolution of protoplanetary disks at millimeter wavelengths is not due to a decline in the disk mass but efficient agglomeration of dust grains from sub-micron to beyond millimeter sizes within a few Myr. That is, the low luminosities at millimeter wavelengths are explained by having most of the dust mass locked up in relatively large particles with little surface area. This agrees with cosmochemical studies of rapid planetesimal formation and is the first step for core accretion formation of the giant planets.

New and expanded facilities, the Atacama Large Millimeter Array (ALMA) and Jansky Very Large Array (JVLA) respectively, will provide the capability to carry out far more sensitive surveys of protoplanetary disks at millimeter through centimeter surveys in the next decade. We will also learn much more about the gas content in the giant planet forming regions of these disks. Ultimately, we will see images of the dust and gas in a representative sample of disks at 10 AU resolution and learn about the radial dependence of their evolution. This opens up the potential for a rich overlap between astronomical and cosmochemical studies of millimeter sized particles in coming years that will inform us about the formation of planetesimals and the first steps toward planets.

***Acknowledgements:*** This work is supported by NASA grant RSA-1369686 and NSF grant AST-0808144. I thank the referees for their detailed comments and Sasha Krot and Gary Huss for their unwavering efforts to strengthen the ties between the astronomers and cosmochemists in Hawaii. Finally, I offer my congratulations to Klaus, whose life story is inspiring, and whose career is spectacular.